\documentclass[twocolumn,trackchanges]{aastex62}

\newcommand{\project}[1]{\textsl{#1}}

\newcommand{\HST}{\project{HST}}

\newcommand{\Kepler}{\project{Kepler}}

\usepackage{amsmath}
\newcommand{\Mod}[1]{\ \mathrm{mod}\ #1}

\submitjournal{ApJL}

\shorttitle{That's No Moon}
\shortauthors{Kreidberg et al.}

\begin{document}

\title{No Evidence for Lunar Transit in New Analysis of \project{Hubble Space Telescope} Observations of the Kepler-1625 System}

\author{Laura Kreidberg}
\affiliation{Harvard-Smithsonian Center for Astrophysics, 60 Garden Street, Cambridge, MA 02138}
\affiliation{Harvard Society of Fellows, 78 Mount Auburn Street, Cambridge, MA 02138}
\author{Rodrigo Luger}
\affiliation{Flatiron Institute, Simons Foundation, 162 Fifth Ave, New York, NY 10010, USA}
\author{Megan Bedell}
\affiliation{Flatiron Institute, Simons Foundation, 162 Fifth Ave, New York, NY 10010, USA}

\begin{abstract}
    Observations of the Kepler-1625 system with the Kepler and Hubble Space Telescopes have suggested the presence of a candidate exomoon, Kepler-1625b I, a Neptune-radius satellite orbiting a long-period Jovian planet. Here we present a new analysis of the Hubble observations, using an independent data reduction pipeline. We find that the transit light curve is well fit with a planet-only model, with a best-fit $\chi^2_\nu$ equal to $1.01$. The addition of a moon does not significantly improve the fit quality. We compare our results directly with the original light curve from \cite{teachey18b}, and find that we obtain a better fit to the data using a model with fewer free parameters (no moon).  We discuss possible sources for the discrepancy in our results, and conclude that the lunar transit signal found by \cite{teachey18b} was likely an artifact of the data reduction.  This finding highlights the need to develop independent pipelines to confirm results that push the limits of measurement precision.
\end{abstract}


\section{Introduction} \label{sec:intro}
Moons are abundant in the Solar System, and provide clues to the formation history, evolution, and even habitability of the planets they orbit. The great scientific potential of moons has prompted extensive search for lunar companions in exoplanetary systems (exomoons), and creative development of new search techniques \citep[e.g.][]{kipping09a, kipping09b, kipping13, simon10, peters13, heller14, noyola14, hippke15, agol15, sengupta16,  vanderburg18}. 

Recently, a potential exomoon candidate was identified in the Kepler-1625 system \citep{teachey18a}. The host planet, Kepler-1625b, has a radius consistent with that of Jupiter and an orbital period of 287 days.  The first evidence for the exomoon candidate, Kepler-1625b I, was based on observations from \Kepler. The light curve showed drops in stellar flux that were attributed to a transiting exomoon; however, later analysis called this result into question, showing that the moon transit features were highly sensitive to the \Kepler\ reduction pipeline and the algorithm used to detrend the data \citep{teachey18b, rodenbeck18}. 

Subsequent follow-up observations with \HST\ revived the possibility of an exomoon in the system based on two factors: a small drop in the system flux after the planet's transit egress and a transit timing variation \citep[][ hereafter TK18]{teachey18b}. The best fit moon had a large radius (comparable to that of Neptune), and if real, is unlike any moon in the Solar System.

Since the primary evidence for the exomoon now rests on the \HST\ transit light curve, in this work we perform an independent reduction and fit to the \HST\ data and compare it to the results from TK18.

%Of the FIXME transits observed by \Kepler\, FIXME what did they show.  

%Recently, analysis by FIXME of \Kepler\ observations of FIXME suggested a moon. Follow-up happened.

%\citep{rodenbeck18} found moon was model dependent

\section{Observations and Data Reduction} \label{sec:data}
The Kepler-1625 system was observed with 26 continuous \HST\ orbits on 28 - 29 October, 2017 (Program GO 15149: PI: A. Teachey). The observations used the Wide Field Camera 3 (WFC3) G141 grism in staring mode, which fixed the spectrum in a constant position on the detector.  At the beginning of the visit, there was a single exposure taken with the F130N filter, which is used to determine the position of the spectral trace. The following exposures used the G141 grism.  See TK18 for additional description of the observation design.
%with the \texttt{SPARS25}, \texttt{NSAMP}=15 readout pattern (exposure time of 290.8 seconds; 9 exposures per orbit). 

We reduced the \HST\ data using custom software developed in \cite{kreidberg14a}.  This software has yielded consistent results with multiple independent pipelines \citep[e.g.][]{knutson14b, spake18}.  We ran our pipeline on the \texttt{flt} data product provided by the Space Telescope Science Institute (STScI).  In keeping with previous WFC3 analysis, we discarded the first orbit of data, where the instrument systematics have larger amplitude. We also discarded exposures taken during the South Atlantic Anomaly passage (exposures 107, 116, 125, and 126).
%The \texttt{flt} files are corrected for dark current, bias, and nonlinearity, and they are cleaned of cosmic ray hits based on a fit to the up-the-ramp samples. 

To begin the data reduction, we fit the centroid of the direct image with a two-dimensional Gaussian. The centroid position determines the position of the spectral trace, which we calculated using the coeffients provided in the configuration file from STScI: \texttt{G141.F130N.V4.32.conf}\footnote{available at \url{http://www.stsci.edu/hst/wfc3/analysis/grism_obs/calibrations/wfc3_g141.html}}.  To process the spectra, we flatfielded the raw data using the spectroscopic flatfield coefficients provided by STScI in \texttt{WFC3.IR.G141.flat.2.fits}, following the instructions in Section 6 of the aXe User Manual\footnote{\url{http://ane-info.stsci.edu/}}.  We then created an extraction box centered on the spectral trace. We varied the height and width of the box in 1-pixel increments to find the window that minimized the root-mean-square (rms) deviation from the best fit to the transit light curve.  The best was $450  < \mathrm{X} < 574$, and $522 < \mathrm{Y} < 536$, where X and Y are physical pixels in the spectral and spatial direction, respectively. 

We reduced the grism exposures with the optimal extraction routine of \cite{horne86}, which minimizes background noise in the extracted spectrum by preferentially weighting pixels that are dominated by the target spectrum.  The inputs for optimal extraction are the background-subtracted data array, and estimate of the error per pixel, an initial guess for the spectrum and its uncertainty, and a mask array for bad pixels.  For the initial guess of the spectrum, we did a simple box extraction (sum over all rows in the extraction window), and assumed the variance was equal to the box-extracted spectrum.  We subtracted the background from the data array as described in \ref{sec:background}. For the error array, we used a quadrature sum of the photon noise (the square root of the pixel counts), the read noise (12 photoelectrons for \texttt{flt} files; WFC3 Data Handbook\footnote{\url{http://www.stsci.edu/hst/wfc3/documents/handbooks/currentDHB/}}), and the error due to background subtraction. The initial pixel mask marked all pixels as good. 

To optimally extract the spectrum,  we first created a smoothed image by median-filtering each row of the data with a 9-pixel-wide window.  We then normalized the smoothed image by dividing each column by its sum, and multiplied it by the best guess spectrum. We compared the smoothed image to the real data and masked outliers in the data that are greater than a threshold $\sigma_\mathrm{cut} = 7.5$. We then recomputed the best guess spectrum with the new mask and the optimal weights from \cite{horne86}. The process is iterated until no outliers greater than the threshold remain.  To create the broadband transit light curve, we sum each optimally extracted spectrum over all wavelengths. 
%This procedure masks any cosmic rays or bad pixels that were missed by the initial \texttt{flt} calibration.  

\begin{figure}
\includegraphics[width = 0.5 \textwidth]{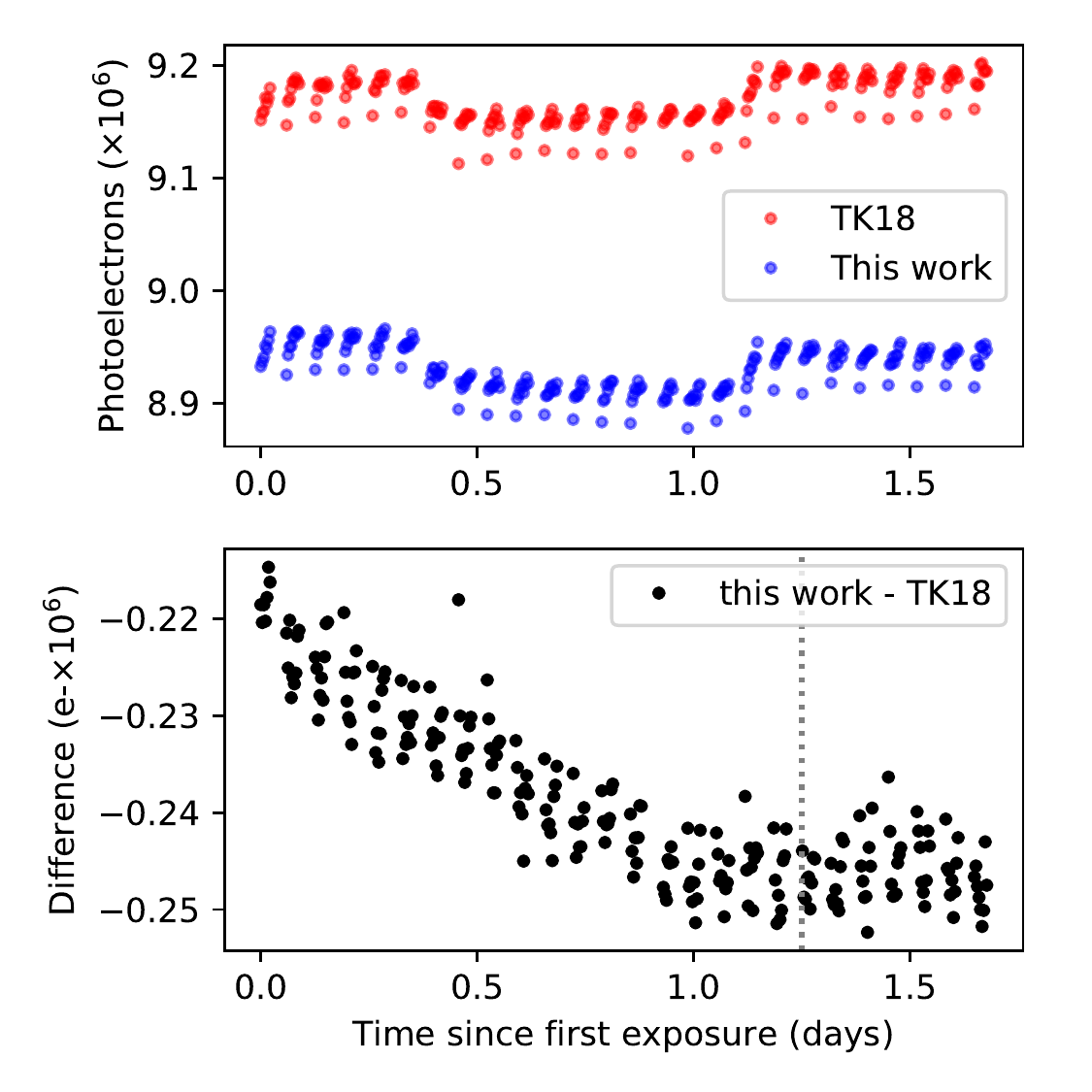}
    \caption{Top: extractred light curve from this work (blue) compared to TK18 (red). Bottom: the difference in photoelectron counts between the data sets. The moon ingress identified by TK18 is marked by the dotted gray line.}
\label{fig:raw}
\end{figure}

The broadband light curve is shown in Figure\,\ref{fig:raw}, in comparison to the light curve from TK18. We note that there are differences between the two data sets, particularly a kink near the moon-like transit feature identified by TK18.

%sky subtraction error - error in mean, plus uncertainty due to extraction box?; uncertainty in mean is very small (0.5 e/pix)
%interpolation to common wavelength scale (row by row) - take out; NO DIFFERNCE
%interpolation based on spectral drift - try both; NO DIFFERENCE
%correct MAD to 1.4826
%change convolution to 4 pix
%convolution for spectral drift -

\subsection{Background Subtraction} 
\label{sec:background} 
The star Kepler-1625 is faint (H mag = 14.0) relative to most other exoplanet host stars observed with WFC3, which makes accurate background subtraction especially important for this target. Moreover, the host star is in a crowded field, so the pixels used to estimate the background must be chosen carefully to avoid contamination from other stars.  To estimate the background counts, we masked pixels with total counts larger than 800 electrons (2.7 electrons/sec) and took the median count in the unmasked pixels. The per pixel uncertainty due to background subtraction is 1.4826 times the median absolute deviation.

%We identified several uncontaminated regions by eye: $130 < \mathrm{X} < 215$ and $6 < \mathrm{Y}  < 24$;  $220 < \mathrm{X} < 250$ and $110 < \mathrm{X} < 155$; $6 < \mathrm{X} < 47$ and $127 < \mathrm{Y} < 141$, where X and Y are pixel numbers in the spectral and spatial direction, respectively (numbering from zero).  

\subsection{Pointing Drift Measurement}
\begin{figure}
\includegraphics[width = 0.5 \textwidth]{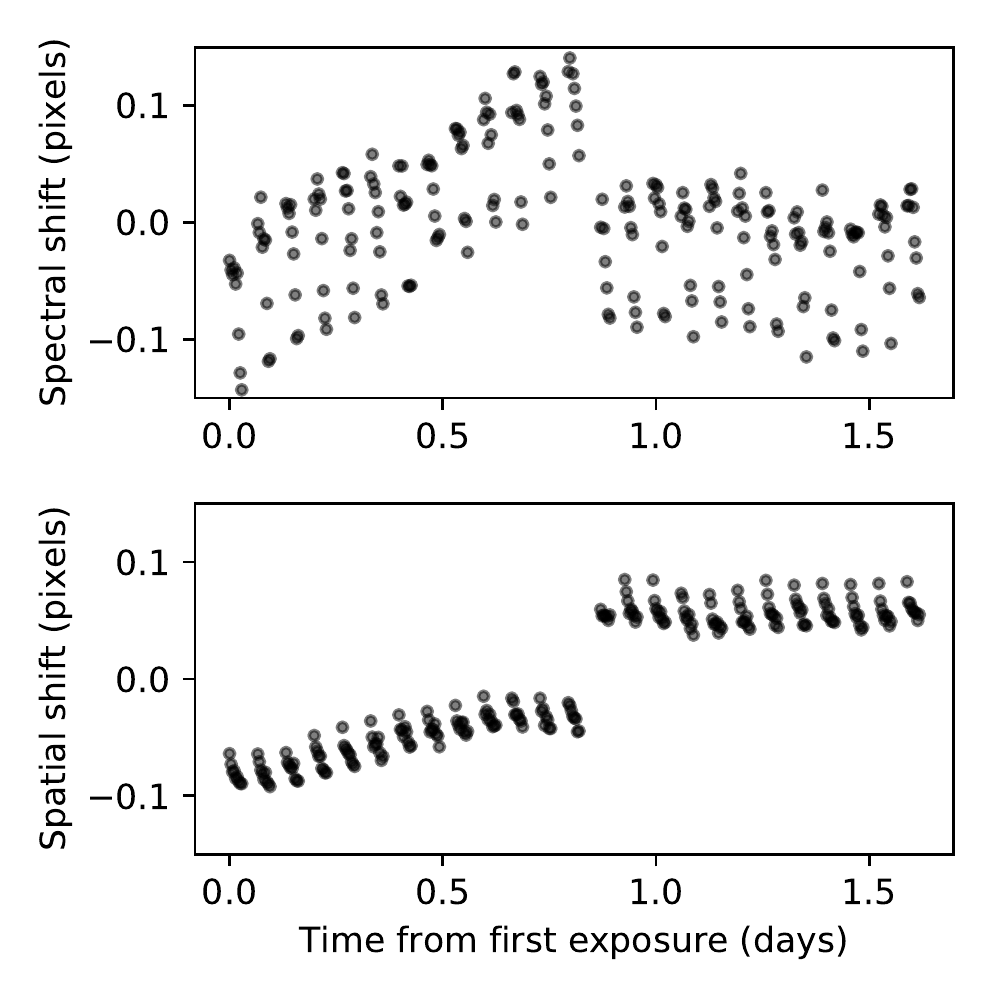}
    \caption{Shift (in pixels) relative to the mean position of the spectrum in the spectral direction (top) and spatial direction (bottom). The largest shift occurs after orbit 14 due to a guide star reacquisition.}
\label{fig:shifts}
\end{figure}

The position of the spectrum on the detector shifts slightly over time ($\sim0.1$ pixel/day) due to the spacecraft's pointing drift. This drift can change the flux measured for the target star: if the spectrum moves onto less sensitive pixels, fewer photoelectrons are recorded. To enable a correction for this effect, we measured the position of the spectrum over time.  Figure\,\ref{fig:shifts} shows the best fit shifts. 

To measure shifts in the spatial direction, we summed each \texttt{flt} image over all columns (which we dub the ``column sum"). We used the first exposure in the visit as a template, and for each subsequent exposure, we used least-squares minimization to calculate the shift in pixels that minimized the difference between its column sum and the template. The shifts are a fraction of a pixel, so we used the NumPy \texttt{interp} routine to do linear interpolation on a sub-pixel scale. The WFC3 point spread function is undersampled, so we convolved each column sum with a 4-pixel-wide Gaussian  before the interpolation \citep[following][]{deming13}.  

To measure the spectral shifts, we repeated this procedure with two differences: (1) we used the optimally extracted spectrum rather than the column sum; and (2) in addition to calculating the best fit shift, we also calculated a best fit normalization factor (a scalar multiple for the whole spectrum), to ensure that our results are not biased by the varying brightness of the host star during the planet's transit. 

%Over the entire 26-orbit visit, the maximum shift is less than 0.2 pixel in the spatial direction and 0.3 pixel in the spectral direction. The largest shift occurs after orbit 14, when the telescope reacquired the guide stars. 

\section{Analysis}
The extracted transit light transit light curve contains both astrophysical signals and instrument systematic noise, which we model simultaneously. 

\subsection{Astrophysics Model}
For the astrophysics, we used the \texttt{planetplanet} package \citep{luger17}, a photodynamical code that calculates light curves for multiple occulting bodies orbiting a star. Within \texttt{planetplanet}, the orbits are computed with the N-body integrator \texttt{REBOUND} \citep{rein12}, which calculates the three-dimensional motion of the star, planet, and moon over time under the influence of gravity. %The \texttt{planetplanet} model returns the system flux at a specified orbital architecture and time. 

In our analysis, we considered two scenarios: a no-moon model and a moon model. The free parameters for the no-moon model were: the stellar radius, the planet radius, the planet's time of central transit, the planet inclination, and the planet mass. For the moon model, we added a third body with six free parameters: moon radius, moon time of central transit, moon orbital period, moon inclination, moon mass, and longitude of the ascending node relative to that of the planet.  We fixed the eccentricity of all bodies to zero. We also fixed the  orbital period of the planet-moon barycenter to 287.378949 days (the best fit from TK18). We elected not to vary the period of the planet-moon barycenter because it is poorly constrained from a single transit observation. A longer period would cause a longer transit duration for the planet, but this could equivalently result from a smaller impact parameter, a larger star, or a massive moon that significantly perturbs the planetary orbit.

We used the following priors: the stellar radius was normally distributed, $R_* \sim N(1.81, 0.17) R_\odot$ (see next section). The planet radius was uniform from $8 - 14R_\oplus$. The planet transit time was uniform over the timespan of the observations, and inclination was uniform from $0 - 90^\circ$.  The planet mass was log-normally distributed, $M_\mathrm{p} \sim 10^{2.5 \pm 0.5} M_\oplus$, based on the expectation for Jupiter-radius objects from \cite{ning18}.  For the moon model, we allowed the transit time to vary uniformly over the entire visit. The moon period was uniform between 1.6 to 260 days. These limits span the duration of the \HST\ observations (so there is one possible moon occultation event), to the orbit at 0.5 the Hill radius, based on the stability limit for prograde moon orbits \citep{domingos06}. The Hill radius calculation assumed the planet and stellar masses are $1\,M_\mathrm{Jup}$ and $1.37\,M_\odot$.  The moon mass varied uniformly from $0 - 30 M_\oplus$. The longitude of the ascending node was also uniform from $0 - 360^\circ$.  The moon inclination was uniform from $0 - 90^\circ$,  and was defined relative to the line of sight. We assigned zero prior probability to scenarios where the moon did not transit or experienced a grazing transit. We made this choice to put an upper limit on the radius of a fully transiting moon; for grazing transits or non-transits the moon radius could be arbitrarily large.

\begin{figure*}
\includegraphics[width = 1.0 \textwidth]{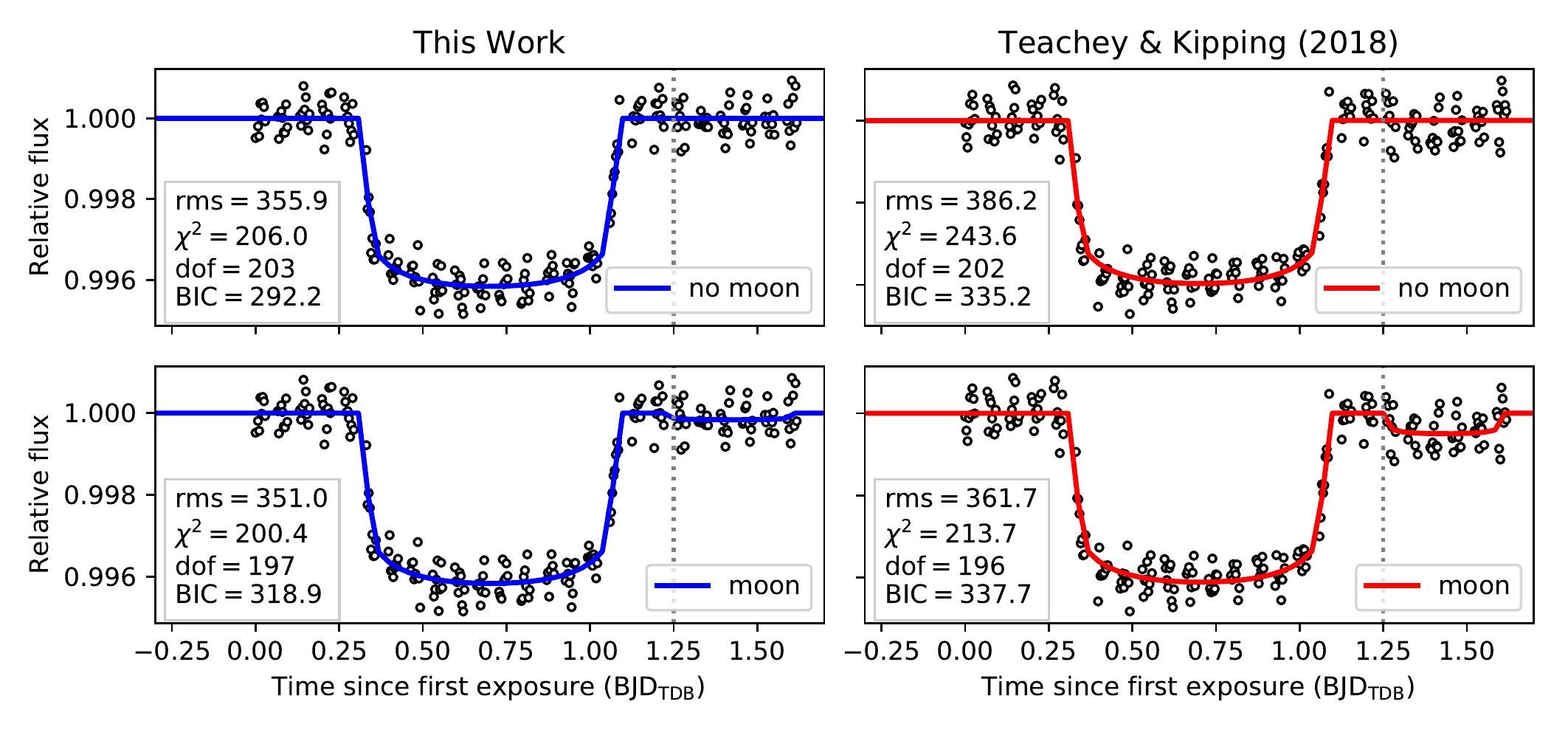}
    \caption{Best fit models compared to normalized transit light curves from this work (left, blue) and from TK18 (right, red). The top panel shows the best fit no-moon model, and the bottom shows the best fit moon model. The lower left of each panel indicates the fit rms (in ppm), the $\chi_2$, the degrees of freedom, and the BIC.  Each light curve is divided by its best fit systematics model (XY decorrelation for this work; exponential and offset for TK18).  The dotted gray line marks the possible moon ingress identified by TK18.}
\label{fig:bestfit}
\end{figure*}

\subsubsection{Stellar Parameters}
For both the moon and no-moon scenarios, we used a quadratic stellar limb darkening law and fixed the coefficients to the prediction for a 5700~K, solar metallicity \texttt{PHOENIX} model from \cite{espinoza15}; $u_1, u_2 = [0.216, 0.183]$.  

We estimatea the host star parameters using the Gaia DR2 parallax \citep{Gaia, GaiaDR2} along with UBV photometry from \citet{Everett2012} and JHK photometry from 2MASS \citep{2MASS}. We employed the isochrone python package \citep{isochrone} with the Dartmouth isochrone grid \citep{Dotter2008} to obtain posterior constraints on the stellar parameters. The resulting parameters indicate that Kepler-1625 has stellar mass $1.37^{+0.13}_{-0.16}$ M$_{\odot}$, radius $1.81^{+0.18}_{-0.16}$ R$_{\odot}$, and age $2.8^{+1.6}_{-1.2}$ Gyr. In our analysis, we fixed the stellar mass to the best fit value ($1.37\,M_\odot$), and used a Gaussian prior on the radius, $R_* \sim N(1.81, 0.17)$.

\subsection{Instrument Systematics Model}
\label{sec:sys}
There are two systematic trends in the data. One is the orbit-long ramp, attributed to charge traps in the detector filling up over the orbit \citep{zhou17}. The other is a visit-long trend over multiple orbits, which could be due to shifts in the target star position onto more/less sensitive pixels.

For our primary fit, we used a linear combination of the spectrum's X and Y position (shown in Figure~\ref{fig:shifts}), multiplied by the non-parametric orbital ramp model from TK18, which assigns each of the nine exposures per orbit a normalization constant, $c_1, ..., c_9$. In sum, for exposure number $i$, the systematics model S is:
\begin{equation}
\label{eq:sys}
    \mathrm{S}_i = c_{j}\times(1 + a\mathrm{X}_i + b\mathrm{Y}_i) 
\end{equation}
where $a$ and $b$, and $c_j$ are free parameters, and $j = i\Mod9 + 1$ is the exposure number relative to the first exposure in the orbit. 

For comparison, we also tested the ``exponential" systematics model from TK18, which combines an exponential visit-long trend, an offset after orbit 14 when the guide stars were reacquired, and the non-parametric orbital ramp model.

\subsection{Light Curve Fits}
We fit the broadband transit light curve using the models described above.  We determined the best fit model parameters with least-squares minimization.  We also ran a Markov chain Monte Carlo (MCMC) fit to determine the posterior probability of the parameters. For the MCMC, we held the ramp parameters $c_1, ..., c_9$ fixed at their best-fit values.  The MCMC used the \texttt{emcee} package \citep{foremanmackey13} with 50 walkers and ran for 5000 steps. We discarded the first 20\% of the MCMC chain as burn-in. As a quick test for convergence, we divided the remainder of the chain in half and confirmed that the results from the first half were consistent with the second half.

\section{Results}
\begin{figure}
\includegraphics[width = 0.5 \textwidth]{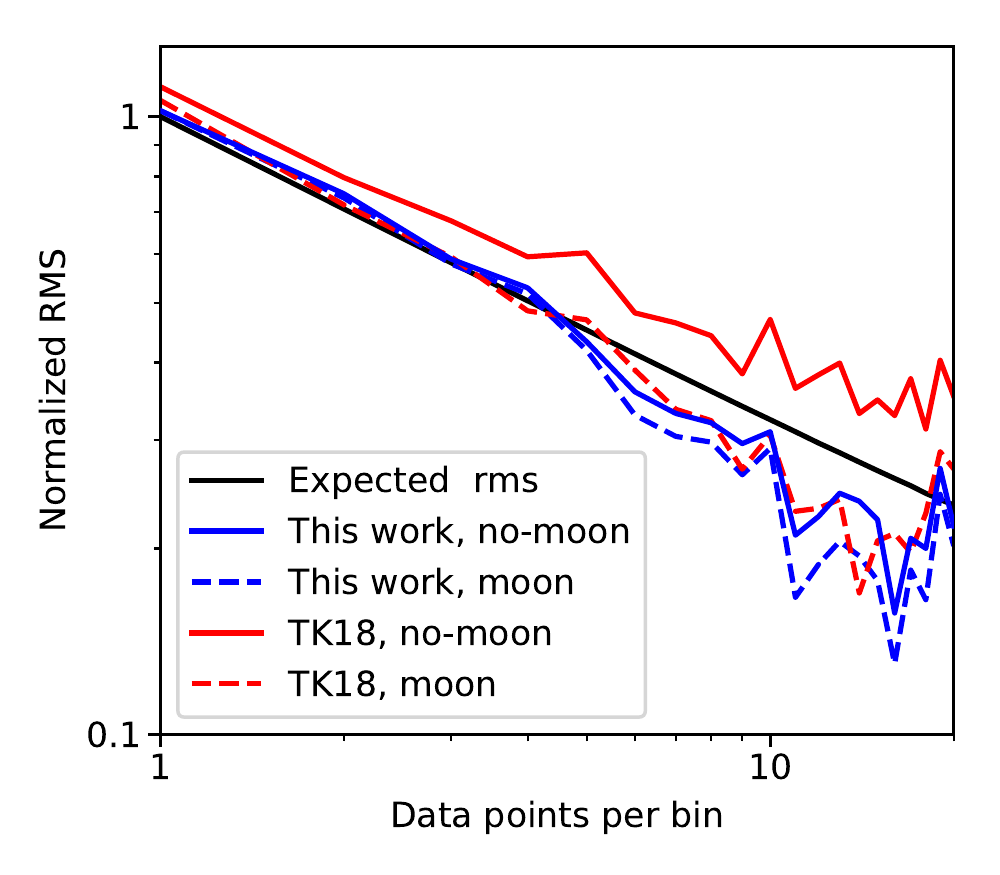}
    \caption{Light curve rms versus bin size for the best fit no-moon model (solid lines) and moon model (dashed lines), for data from this work (red) and TK18 (blue). The fits to data from this work agree well with the expected photon-limited, $\sqrt{N}$ decrease in rms with bin size (black line).  We also reach the photon limit for the TK18 data, but only for the moon model. The rms for the no-moon fit (red dashed line) ranges from $1.1 - 1.5\times$ the photon limit for 1 to 20 points per bin.}
 \label{fig:rms}
\end{figure}

We obtained an excellent fit to the light curve with the no-moon model, as illustrated in  Figure\,\ref{fig:bestfit}. The residuals to the no-moon model fit have rms equal to 356 parts per million (ppm), which is within 3\% of the predicted photon shot noise (367 ppm), and yields a $\chi_\nu^2 = 1.01$. The binned rms decreases with the square root of the number of points per bin, as expected for photon noise-limited statistics (see rms versus bin size in Figure~\ref{fig:rms}).

We obtained a slightly lower rms with the moon model (351 ppm); however, this is not a large enough improvement in fit quality to merit the addition of six additional free parameters.  The moon model has a small increase in $\chi^2_\nu$ to 1.02. According to the Bayesian information criterion (BIC), which penalizes unnecessary model complexity, the moon model is disfavored with $\Delta\mathrm{BIC} = 26.7$. This constitutes strong evidence against the inclusion of a moon \citep{kass95}.  In addition, as shown in Figure\,\ref{fig:corner}, the posterior distribution for the moon transit time spans the entire duration of the observations ($2\sigma$ confidence).  The upper limit on the moon radius is $3.6\,R_\oplus$ at $3\sigma$ confidence.

%The best fit moon model has a radius $r_\mathrm{moon} = 2.4R_\oplus$, a mid-transit time $t_\mathrm{moon} = 2458055.2869\,\mathrm{BJD_{TDB}}$, and a period $P_m = 26.5$ days. This fit has a slightly lower rms than the no-moon model (359.5 ppm versus 362.2 ppm), but 

We found that our results were unchanged when we used the exponential systematics model from TK18 (described in \S\,\ref{sec:sys}). For this case, the moon model is also strongly disfavored ($\Delta\rm{BIC} = 32.2$). The fit rms is within 5 ppm of the XY decorrelation model.

The posterior probabilites for the planet's mass and radius are consistent with either a gas giant planet or brown dwarf. The radius is $11.1^{+0.5}_{-0.2}\,R_\oplus$ and mass is $10^{2.2\pm0.5}\, M_\oplus$.

\subsection{Comparison with Teachey \& Kipping (2018)}
TK18 found evidence for the transit of a Neptune-size moon in their analysis of the HST data, in contrast to the findings presented here.  To compare our results with theirs, we fit the TK18 light curve directly.  We fit the astrophysical signal with both the no-moon and models, and used the exponential systematics model.  Figure\,\ref{fig:bestfit} shows the best fit models. 

%For the systematics, we tested the XY decorrelation model described above, but found that it did not perform as well as the systematics models presented in TK18 (10\% higher rms and time-correlated residuals).  We therefore opted to use a TK18 systematics model to enable a fair comparison between our results. We obtained the best fit using an exponential trend in time, a constant offset after orbit 14 (where the guide stars are reacquired), and the non-parameteric orbit-long ramp model.  This model has one more free parameter than the XY decorrelation model given in Equation~\ref{eq:sys}. 

Similar to the findings of TK18, the moon model improved the fit quality by a $\Delta\chi^2 = 29.9$. Notably, however, the moon model fit to the TK18 data does not perform better than the no-moon model fit to our new data (rms of 362 versus 356 ppm), even with the additional seven free parameters.  

\begin{figure}
\includegraphics[width = 0.5 \textwidth]{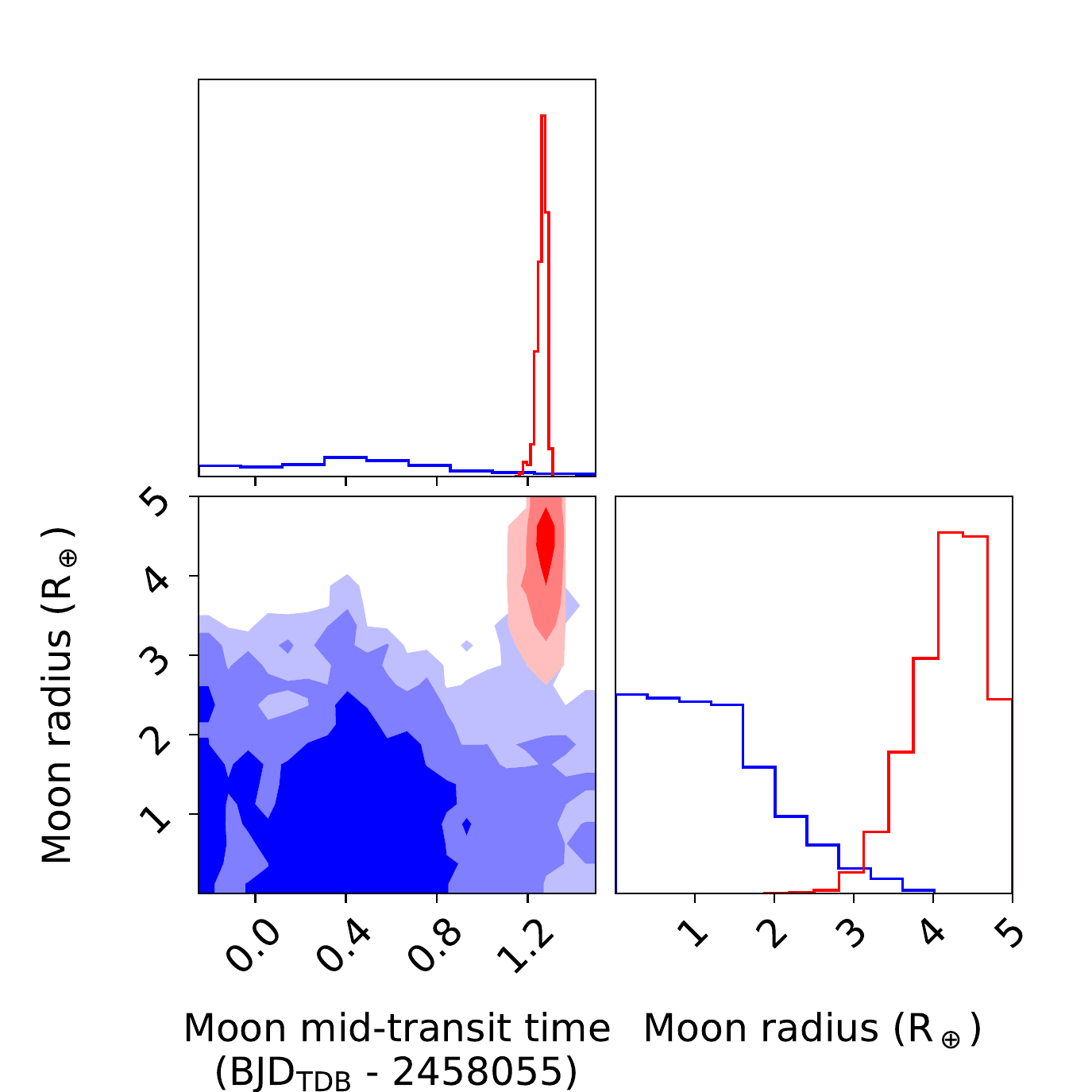}
    \caption{Posterior distributions for the moon radius and time of transit based on an MCMC fit to data from this work (blue) and from TK18 (red). The shading corresponds to $1-$, $2-$, and $3\,\sigma$ contours (from darkest to lightest). These values are marginalized over all other model parameters.} 
\label{fig:corner}
\end{figure}

The moon model also yields qualitatively different posterior distributions for the two data sets.  As shown in Figure\,\ref{fig:corner}, for the TK18 data the moon radius and transit time are peaked at $r_\mathrm{moon} = 4.3\pm0.5\,R_\oplus$  and $t_\mathrm{moon} =  2458056.29^{+0.06}_{-0.04}\,\mathrm{BJD_{TDB}}$. By contrast, the fit to the new data presented here yields an upper limit on the moon radius of $3.6\,R_\oplus$ at $3\sigma$ confidence, and the transit time is unconstrained.

Although the two data sets yield different constraints on the moon properties, the planet's mid-transit time agrees to better than $1\sigma$ for the two fits. The transit time is earlier than expected based on the Kepler data ($3\sigma$ confidence; TK18), suggesting that there are transit timing variations in the system. Such variation could arise from the presence of a moon, as suggested by TK18; however, the variation could also be caused by another planet in the system.

%, finding $t_\mathrm{planet} = 2458055.5545 \pm 0.0013$ versus $2458055.5563 \pm 0.0014\,\mathrm{BJD_{TDB}}$ from TK18.  This result leaves open the possibility for transit timing variations in the system.  

\section{Discussion \& Conclusions}
A natural question arising from our analysis is the source of the discrepancy between TK18 and the new results presented here. We find that with our new data, there is strong evidence against the moon ($\Delta\rm{BIC} > 25$), even when we use a comparable model to TK18 (a 6-parameter moon model and an exponential fit to the visit-long trend).

If the model is not the source of the difference, it must be the extracted transit light curves themselves.  Figure\,\ref{fig:raw} shows a direct comparison of the light curves. The count rate we measure is $2.46 -2.74\%$ lower than the TK18 light curve, and there is a small bump in the difference between the two data sets near the location of the moon transit identified in the TK18 data (see the bottom panel). This bump may be the source of the moon feature reported in TK18.

 We explored several modifications to our pipeline to attempt to reproduce the TK18 data reduction. These included rotating the image by 0.5 degrees, using the same aperture as TK18 to extract the spectrum, and scaling the master sky flat for the background subtraction (rather than just subtracting the median). None of these modifications had a significant effect on our results. 

There are a few other steps in the TK18 data reduction that would require substantial modification of our pipeline to recreate, but seem unlikely to be responsible for the difference. One of these is outlier masking. TK18 identify outliers with a Gaussian process fit to the pixel-level light curves, compared to our optimal extraction approach. Despite the difference, both methods flag $\sim$0.01\% of pixels as bad.  We also do not use the STScI software \texttt{aXeprep} to embed the raw $256\times256$ image in a larger array; however, this process primarily affects the edge of the image, many pixels distant from the extraction box, so it is unclear how this step would bias the light curve.  We conclude that no single choice in the data reduction provides an easy explanation for the difference in our light curves. 

During the referee process for this work, we learned of another manuscript that also reanalyzed the \HST\ transit observation \citep{heller19}. The best fit favored a moon model similar to that found by TK18;  however, an MCMC analysis did not converge on this model, leading the authors to conclude that the highest likelihood solution may be an outlier.
 
Taken together, these findings illustrate the challenge of pushing measurement precision to the 100 ppm level, and highlight the importance of developing multiple independent pipelines to confirm cutting-edge results.

\acknowledgments
We thank A. Teachey for helpful discussions and for providing the extracted light curve from TK18.  We also thank the anonymous referee for a thoughtful report that improved the manuscript. The HST data presented in this paper were obtained from the Mikulski Archive for Space Telescopes (MAST). STScI is operated by the Association of Universities for Research in Astronomy, Inc., under NASA contract NAS5-26555. Support for MAST for non-HST data is provided by the NASA Office of Space Science via grant NNX13AC07G and by other grants and contracts.  We also use data from the European Space Agency (ESA) mission {\it Gaia} (\url{https://www.cosmos.esa.int/gaia}), processed by the {\it Gaia} Data Processing and Analysis Consortium (DPAC, \url{https://www.cosmos.esa.int/web/gaia/dpac/consortium}). Funding for the DPAC has been provided by national institutions, in particular the institutions participating in the {\it Gaia} Multilateral Agreement.  
%We also use data products from the Two Micron All Sky Survey, which is a joint project of the University of Massachusetts and the Infrared Processing and Analysis Center/California Institute of Technology, funded by the National Aeronautics and Space Administration and the National Science Foundation.  This research has made use of the SIMBAD database, operated at CDS, Strasbourg, France; the REBOUND integrator package \citep{rein12}; the NumPy package \citep{van2011numpy}; and NASA's Astrophysics Data System. 

\bibliographystyle{aasjournal}
\bibliography{ms.bib}

\end{document}